\documentclass[reprint,superscriptaddress,amsmath,amssymb,aps,pra,]{revtex4-2}

\usepackage{graphicx}
\usepackage{dcolumn}
\usepackage{bm}
\usepackage{mathtools}
\usepackage{hyperref}

\usepackage{graphicx, color}
\usepackage{braket}
\usepackage [english]{babel}
\usepackage [autostyle, english = american]{csquotes}
\MakeOuterQuote{"}

\definecolor{chromeyellow}{rgb}{1.0, 0.65, 0.0}

\usepackage{color}
\usepackage[dvipsnames]{xcolor}
\usepackage[normalem]{ulem}

\newcommand{\init}{{\rm i}} 
\newcommand{\final}{{\rm f}} 
\newcommand{\indist}{{\rm ind}} 
\newcommand{\dist}{{\rm dis}} 
\newcommand*{\expval}[1]{\left\langle  #1  \right\rangle}

\usepackage[title,titletoc]{appendix}
\usepackage{array}
\usepackage{placeins}

\begin{document}

\title{Bosonic statistics enhance Maxwell's demon in photonic experiment}

\author{Malaquias Correa Anguita}
\affiliation{MESA+ Institute for Nanotechnology, University of Twente, P.O.~Box 217, 7500 AE Enschede, Netherlands}

\author{Sara Marzban}
\affiliation{MESA+ Institute for Nanotechnology, University of Twente, P.O.~Box 217, 7500 AE Enschede, Netherlands}

\author{William F. Braasch Jr.}
\affiliation{Joint Center for Quantum Information and Computer Science, NIST and University of Maryland, College Park, MD 20742, USA}

\author{Twesh Upadhyaya}
\affiliation{Joint Center for Quantum Information and Computer Science, NIST and University of Maryland, College Park, MD 20742, USA}
\affiliation{Department of Physics,
University of Maryland, College Park, MD 20742, USA}

\author{Gabriel Landi}
\affiliation{Department of Physics and Astronomy, University of Rochester, Rochester, New York 14627, USA}

\author{Nicole Yunger Halpern}
\affiliation{Joint Center for Quantum Information and Computer Science, NIST and University of Maryland, College Park, MD 20742, USA}
\affiliation{Institute for Physical Science and Technology, University of Maryland, College Park, Maryland 20742, USA}

\author{Jelmer J. Renema}
\affiliation{MESA+ Institute for Nanotechnology, University of Twente, P.O.~Box 217, 7500 AE Enschede, Netherlands}

\date{\today}

\begin{abstract}
Maxwell’s demon elucidates the value of information in thermodynamics, using measurement and feedback: he evolves an equilibrated gas into a nonequilibrium state, from which one might extract work. The demon can evolve the system farther from equilibrium,
on average, if the particles obey Bose–Einstein statistics than if they are distinguishable. We experimentally support
this decade-and-a-half-old prediction by comparing indistinguishable with distinguishable photons. We use a fully programmable linear-optics platform, whose local photon statistics were shown recently to behave thermally.
Our demon nondestructively measures the number of photons in a subset of the modes. Guided by the outcome, he conditionally interchanges the measured and unmeasured modes. This interchange creates a positive temperature difference between a mode in a particular subset and a mode in the other. The temperature difference is greater, on average, if the photons are indistinguishable. This result bolsters a long-standing prediction about the interplay among thermodynamics, information, and quantum particle statistics. Additionally, this work suggests a thermodynamic means of weakly validating boson-sampling platforms.
\end{abstract}

\maketitle

Much of modern thermodynamics revolves around the question \emph{how can information and quantum phenomena benefit thermodynamic tasks~\cite{Goold_16_Role,Vinjanampathy_16_Quantum,Deffner_19_Quantum}?} A Gedankenexperiment of Maxwell's, and extensions thereof, epitomize this investigation~\cite{Leff_90_Maxwell,Leff_02_Maxwell,Maruyama_09_Physics}: Maxwell introduced a being, later called a demon, that can measure particles' velocities. Using the outcomes, the demon creates a temperature gradient from a uniform thermal state. Szilard recast the thought experiment in terms of a particle-number gradient~\cite{Szilard_29_Uber}. Kim \emph{et al.} quantized Szilard's argument by 2011~\cite{Kim_11_Quantum}. They envisioned a two-particle gas in a box halved by a piston. Consider measuring the particles' locations. Afterward, one can conditionally interchange the box's halves, so that a particular side contains most of the particles. The greater the imbalance, the farther the system from equilibrium. Bosons bunch together more than distinguishable particles or fermions, on average. Particle statistics thereby affect how far-from-equilibrium a state one can prepare, on average, using information. This theoretical result has remained untested until now. We support the prediction with a linear-quantum-optics experiment, as well as extending prior calculations analytically.

Linear quantum optics is a quantum-information-processing paradigm centered on noninteracting bosons. The formalism's linearity precludes interactions. However, bosons' indistinguishability enables interference, which enables a quantum advantage under reasonable complexity-theoretic assumptions~\cite{Gard_15_Introduction,Brod_19_Photonic, Aaronson_11_Computational}.

\emph{Boson sampling} epitomizes this advantage~\cite{Aaronson_11_Computational}. To define boson sampling, we imagine preparing one photon or a vacuum in each of several optical modes. Suppose this system undergoes a linear-optical transformation selected according to the Haar measure (uniformly randomly, loosely speaking). Consider measuring, afterward, the number of photons per mode. The outcome is distributed according to some probability distribution. Sampling from that distribution, one performs boson sampling. Classical computers cannot perform this task efficiently, under reasonable complexity-theoretic conjectures. This result has spurred experiments, including claimed demonstrations of quantum advantage~\cite{Broome_2013_sampling, Spring_2013_sampling, Tillmann_2013_sampling, Crespi_2013_sampling, PanSampling, zhong_quantum_2020, zhong_phase-programmable_2021, Deng_sampling_2023}. 

Researchers have recently uncovered thermodynamic properties of linear quantum optics~\footnote{
This work complements two other types of bosonic thermodynamics. One, familiar from textbooks, includes blackbody radiation and  lasers. Unlike these phenomena, linear quantum optics involves all-to-all connectivity among equal-energy modes, is globally far from equilibrium, and conserves photon number. Second, more-recent quantum-optical thermodynamics does not meet these conditions or does not exhibit many-photon interference~\cite{13_Martinez_Dynamics,19_Zanin_Experimental,23_Aifer_Thermodynamics}.}. 
Consider, again, the boson-sampling procedure. After the unitary, a few modes are approximately in an equilibrium state, although the whole-system state is pure~\cite{somhorst_quantum_2023}. One can thereby define a linear-optical system's temperature \cite{Somhorst_thesis2021}. These results may be surprising, since equilibration often results from interactions~\cite{92_Goldenfeld_Lectures}, which linear-optical bosons do not undergo. To what extent can linear quantum optics exhibit thermodynamic features such as heat, work, engines, and the interplay between energy and information?

We expand linear optics' nascent quantum thermodynamics to a Maxwell demon augmented by spin statistics. Our experiment featured a boson-sampling system. After performing the Haar-random unitary, we partitioned the modes into two equal-size subsets. We measured which subset contained more photons. We interchanged the sets, conditionally on the outcome, during postprocessing. Consequently, a preselected set ended up with more photons. A nonequilibrium state resulted from this measurement and feedback. The photons were indistinguishable in one batch of trials and distinguishable in another. Our experiment demonstrates that a Maxwell demon can produce a greater particle-number and temperature difference, on average, using indistinguishable particles than using distinguishable ones. Hence we experimentally support a prediction in~\cite{Kim_11_Quantum}. Our work extends earlier photonic Maxwell demons, which amplify classical fluctuations~\cite{Vidrighin2016Photonic, Zanin2022enhancedphotonic}: quantum fluctuations dominate in our experiment, since our quantum system is approximately closed \cite{somhorst_quantum_2023}. 

The rest of the paper is organized as follows. We review Maxwell's demon, then thermodynamics of linear quantum optics. Next, we describe the experimental setup and results. Their significance, and the opportunities they engender, follow.

\emph{Background: Maxwell's demon.---}Maxwell envisioned a being 
who steers a thermal classical gas into a nonequilibrium macrostate, apparently contradicting the second law of thermodynamics~\cite{maxwell1871theory}. Consider a box containing a thermally equilibrated gas.
The ``demon''~\cite{Thomson1874Kinetic} partitions the box into halves $A$ and $B$, using a wall that contains a door.
Suppose a molecule approaches $A$ from $B$. If the molecule's kinetic energy exceeds the average in $A$, the demon opens the door. Conversely, he filters low-energy molecules into $B$. His actions need not cost work, Maxwell argued. 
%
Heating up the warm $A$, the demon seemingly violates the second law. Szilard realized that the demon could generate, rather than a temperature gradient, a particle-number gradient~\cite{Szilard_29_Uber}. 

The latter gradient depends on quantum particle statistics, Kim \textit{et al.} realized~\cite{Kim_11_Quantum}.
Their demon manipulates a two-particle gas, measuring whether the particles are in $A$ or $B$. Afterward, he can interchange the box's halves, so more particles end up in a chosen side. Bosons tend to bunch, so they create the greatest particle-number gradient, on average. Fermions spread out, by Pauli's exclusion principle, producing the smallest average gradient. Distinguishable particles interpolate between the extremes. 

Our work build on~\cite{Kim_11_Quantum} in multiple ways. Kim \emph{et al.} model two-particle gases via high-level, abstract theory. We present an experiment. We also extend calculations of theirs to arbitrarily many particles, in~\cite{SuppMat}. Additionally, we stipulate how to prepare the gases in the necessary initial states, leveraging the following results.

 \begin{figure}
     \centering
     \includegraphics[width = 0.47\textwidth]{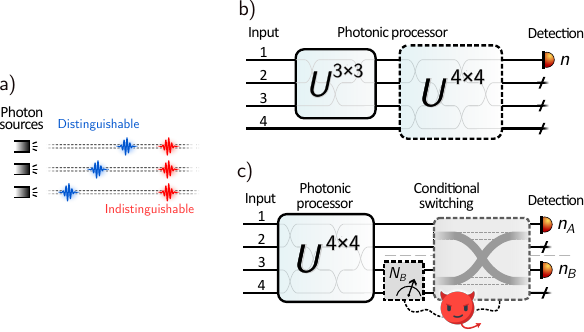}
     \caption{Experimental schematic.
    (a) State preparation.
    (b) Composite-system equilibration.
    (c) Maxwell-demon experiment.
    The dashed horizontal line partitions the modes into subsets $A$ (top) and $B$ (bottom).
    See the main text for details.
    }
\label{fig:Fig2_ExpSetup}
\end{figure}

\emph{Background: quantum thermodynamics of linear optics.---}We briefly review the relevant features of linear quantum optics~\cite{loudon_multimode_2000}. 
During the prototypical linear-quantum-optics experiment, we prepare at most one photon per optical mode. The state evolves under a photon-number-conserving interferometer represented by a unitary $U$. $U$ mixes the modes, interfering the photons. The evolution produces a superposition of Fock states. We measure the state in the Fock basis. The outcome is a list of numbers, each specifying the number of photons in one mode. The probability distribution over the possible outcomes is the \emph{outcome distribution}.

The connection between linear quantum optics and thermodynamics arises from~\cite{Aaronson_11_Computational,arkhipov2012bosonic}. Consider averaging the output state over infinitely many Haar-random unitaries. The result is the maximally mixed state of $N$ photons in $M$ modes, regardless of the input~\cite{Aaronson_11_Computational,arkhipov2012bosonic}.
 
Building on this result, Somhorst \textit{et al.} showed experimentally that Haar-averaging linear-optical dynamics leads to local thermal behavior~\cite{somhorst_quantum_2023,Somhorst_thesis2021}.
Consider tracing out everything except one mode from the maximally mixed $N$-photon state. The remaining mode exhibits canonical, thermal-light statistics. So does one output mode if the photonic system undergoes linear-optical dynamics, on average with respect to the Haar measure, by~\cite{Aaronson_11_Computational,arkhipov2012bosonic}.
The local thermal state's effective temperature depends on the photon energy $E_\nu$, Boltzmann’s constant $k_\mathrm{B}$, and the photon density $N/M$~\cite{Somhorst_thesis2021}:
\begin{equation}
\label{eq:somhorsttemperature}
    T = \frac{E_\nu}{k_\mathrm{B}}  \, 
    \frac{1}{\ln(1+[N/M]^{-1})} \, .
\end{equation}
A near-infrared photon's $E_\nu / k_{\rm B} \approx 10^3$ to $10^4$.

\emph{Experimental setup.---}Our experiments involved the following components. Three sources produced one photon apiece via parameteric down-conversion.
We injected the photons into an $M$-mode reprogrammable silicon-nitride photonic processor~\cite{TriPlex, QuiX2021,nist_note}, such that each mode contained one photon or a vacuum. The processor implemented a unitary. Using superconducting-nanowire single-photon detectors \cite{ReviewSNSPD, Marsili-SNSPD, SuppMat}, we counted the photons in each output mode relevant to the experiment performed.

We modulated the photons' indistinguishability by controlling their times of entry into the photonic chip [Fig.~\ref{fig:Fig2_ExpSetup}(a)]. If not delayed, the photons exhibited maximal indistinguishability and approximate bosonic statistics. If delayed sufficiently, they became distinguishable and obeyed classical statistics~\cite{SuppMat}.

\emph{Composite-system equilibration.---}To showcase the setup, we invert the demon's action (before emulating him in another experiment). He separates an equilibrated system into two different-temperature systems; we equilibrate two different-temperature systems with each other. Reference~\cite{arkhipov2012bosonic} predicts this equilibration.

Figure~\ref{fig:Fig2_ExpSetup}(a,b) illustrates the process. We conducted (i) experiments with distinguishable photons and (ii) otherwise identical experiments with indistinguishable photons. In each experiment, we injected $N{=}3$ photons into an $M{=}3$ interferometer that implemented a Haar-random unitary $U^{3\times3}$. Afterward, we measured the number of photons in the first output mode (chosen without loss of generality, since all output modes have identical Haar-averaged statistics). We performed this protocol with each of 50 Haar-random unitaries. 

In an extended protocol, an $M{=}4$ interferometer followed the $M{=}3$ one. The second interferometer applied a Haar-random unitary $U^{4\times4}$ [dashed outline in Fig.~\ref{fig:Fig2_ExpSetup}(b)]. This process equilibrated the first interferometer's output with a vacuum mode---a canonical thermal state with $N/M=0$ and hence $T=0$ [Eq.~\eqref{eq:somhorsttemperature}]. Afterward, we measured the number of photons in the first mode. We performed this extended process with 50 unitary pairs.

\begin{figure}
    \includegraphics[width=0.48\textwidth]{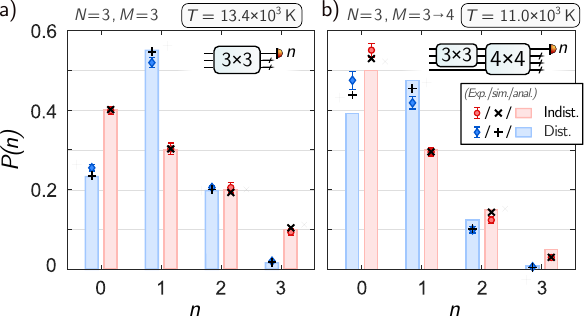}
    \caption{\label{fig:GasEquilibration}
    Composite-system equilibration.
    Probability $P(n)$ of detecting $n$ photons in one mode, averaged over Haar-random unitaries.
    (a) $M{=}3$ experiment.
    (b) Extended equilibration experiment featuring $M{=}3$ and $M{=}4$ interferometers.
    Insets depict photonic-processor configurations.
    See the main text for details.
    }
\end{figure}

Figure~\ref{fig:GasEquilibration} shows the results. Denote by $P(n)$ the probability of detecting $n$ photons in the measured mode.
Figure~\ref{fig:GasEquilibration}(a) shows the $M {=} 3$ experiment's $P(n)$, averaged over the 50 unitaries. Figure~\ref{fig:GasEquilibration}(b) shows the second experiment's $P(n)$, averaged over the 50 unitary pairs. Indistinguishable photons produced the red circles; and distinguishable photons, the blue diamonds.
Error bars represent standard deviations. We produced the crosses by numerically simulating indistinguishable photons; and the plus signs, distinguishable photons. We averaged the numerical results over the Haar-random unitaries performed experimentally. The shaded bars follow from analytical calculations based on symmetry properties of the Haar ensemble~\cite{SuppMat}. 

The experimental, numerical, and analytical results largely agree. Disagreements are largest on the left-hand side of Fig.~\ref{fig:GasEquilibration}(b). Experimental results deviate from numerics due to imperfections in the implemented unitaries. Numerics differ from analytics due to the finite sampling of Haar-random matrices. 

Let us compare the local temperatures achieved before and after the final interferometer. 
Denote by $P_{\mathrm{exp}}(n)$ the empirically observed probability of detecting $n$ photons in the measured mode.
To calculate the local temperature, we compared $P_{\mathrm{exp}}(n)$ with a parameterized distribution $P_{\mathrm{fit}}(n;N,M)$. We identified the $N$ and $M$ values that minimized the total variation distance. Substituting these values into Eq.~\eqref{eq:somhorsttemperature} yielded the local temperature~\cite{Somhorst_thesis2021}.
After the $M{=}3$ interferometer, mode 1 achieves the temperature $12.96 \pm 1.91\ \mathrm{kK}$, in agreement with the $N/M {=} 1$ prediction $T=13.39\ \mathrm{kK}$ from Eq.~\eqref{eq:somhorsttemperature}. 
After the $M{=}4$ interferometer, the local temperature is $9.72 \pm 0.77\ \mathrm{kK}$. It lies near the $N/M {=} 3/4$ prediction $T=10.96\ \mathrm{kK}$. These temperatures should be interpreted as energy scales set by the photon energy $E_\nu$; for near-infrared photons, $E_\nu/k_\mathrm{B}$ is on the order of $10^4\ \mathrm{K}$, consistent with the values shown. These results illustrate that each unitary equilibrates the system locally, irrespectively of the input.

\begin{figure*}
    \includegraphics[width=0.97\textwidth]{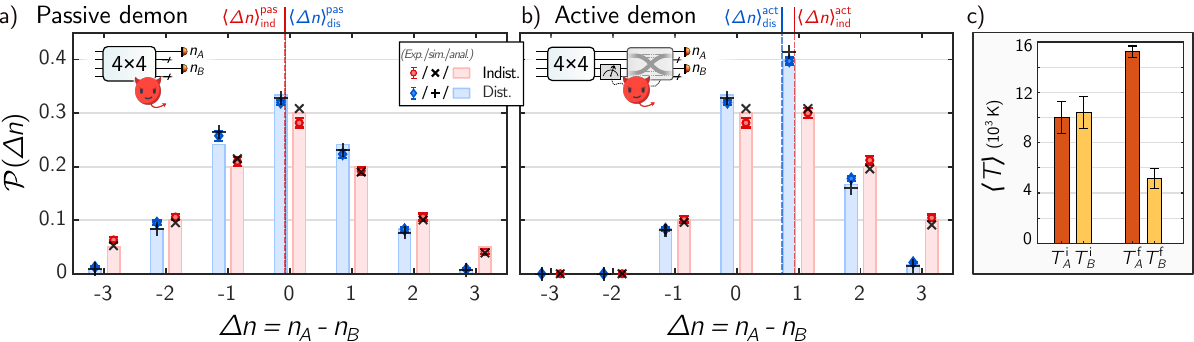}
    \caption{\label{fig:PhNdiff}
    Photonic Maxwell-demon experiment.
    Probability distributions $\mathcal{P}(\Delta n)$ over the photon-number difference $\Delta n \coloneqq n_A - n_B$, averaged over Haar-random unitaries.
    Insets depict photonic-processor configurations.
    (a) Passive-demon experiment.
    (b) Active-demon experiment.
    (c) Average local temperatures $\langle T \rangle$ in the active-demon experiment. 
    See the main text for details.
    }
\end{figure*}

\emph{Photonic Maxwell demon enhanced by particle-number statistics.---}Having demonstrated our experimental setup, we exhibit a linear-quantum-optical Maxwell demon. Figure \ref{fig:Fig2_ExpSetup}(a,c) shows a schematic. The photons were indistinguishable in one batch of trials and distinguishable in another. We prepared $N{=}3$ photons so that each mode contained one photon or a vacuum. An $M{=}4$ interferometer evolved the photons under a Haar-random unitary $U^{4\times4}$. Then, we partitioned the modes into two subsets: $A$ (the top two modes) and $B$ (the bottom two). Each mode behaved thermally, according to the previous section. We observed this thermality by measuring the number of photons in each subset's topmost mode. Ideally, we sacrificed no generality by choosing this partition and the top mode: the Haar-averaged outcome distribution is invariant under mode permutations. Experimental imperfections, and the finite number of Haar-random matrices used, break this symmetry weakly, without affecting the conclusions~\cite{SuppMat}.

In protocol 1, we acted as a passive demon, performing no measurement-and-feedback protocol. We only measured (i) the number $n_A$ of photons in subset $A$'s top mode and (ii) the analogous $n_B$. Define the photon-number difference $\Delta n \coloneqq n_A - n_B$. We performed this protocol with each of 100 Haar-random unitaries.

In protocol 2, we simulated an active demon. After $U^{4\times4}$, effectively, we measured the number $N_B$ of photons in subset $B$. (We performed this step by classically postprocessing the detection record. The step was equivalent to a quantum nondemolition measurement.) The ``measurement'' revealed nothing about the photons' distribution within $B$. If $N_B > N/2$, we interchanged the two subsets [intertwined gray curves in Fig.~\ref{fig:Fig2_ExpSetup}(c)]. Otherwise, we did not (thin, dashed lines). Finally, we measured $n_A$ (the number of photons in $A$'s topmost mode) and $n_B$. We performed this protocol with each of the Haar-random unitaries used in protocol 1. 

Figure~\ref{fig:PhNdiff} presents the average-over-unitaries probability distributions $\mathcal{P}(\Delta n)$ over the possible photon-number differences. Figure~\ref{fig:PhNdiff}(a) shows the passive demon's $\mathcal{P}(\Delta n)$; and Fig.~\ref{fig:PhNdiff}(b), the active demon's. Indistinguishable photons led to the red disks; and distinguishable particles, the blue diamonds.
Error bars indicate standard deviations. Numerical simulations, represented by plus signs and crosses, featured the same unitaries. The shaded bars represent analytical predictions~\cite{SuppMat}. According to Fig.~\ref{fig:PhNdiff}(a) the passive demon's $\mathcal{P}(\Delta n)$ is symmetric about $\Delta n = 0$, as expected from the Haar ensemble's permutation invariance (invariance under permutations of the modes). The active demon's distribution has more weight on high $\Delta n$ values [Fig.~\ref{fig:PhNdiff}(b)]. We elucidate this asymmetry's implications below.

The average photon-number difference $\langle \Delta n \rangle$ indicates the final state's distance from internal equilibrium. In the passive-demon experiment [Fig.~\ref{fig:PhNdiff}(a)], indistinguishable particles led to an average photon-number difference
$\langle \Delta n\rangle ^\mathrm{pas}_\mathrm{ind} 
= -0.082 \pm 0.029$; 
and distinguishable particles, to $\langle \Delta n\rangle ^\mathrm{pas}_\mathrm{dis} 
=-0.070 \pm 0.018$. 
Both averages vanish ideally, by the Haar average's permutation invariance. The averages deviate slightly from zero because we sampled finitely many unitaries. 
The active-demon experiment led to greater $\expval{\Delta n}$ values, because its distributions $P(n)$ were more lopsided [Fig.~\ref{fig:PhNdiff}(b)]. Indistinguishable particles led to an average photon-number difference $\langle\Delta n \rangle ^\mathrm{act}_\mathrm{ind} 
= 0.937 \pm 0.027$; and distinguishable photons, to the lesser $\langle\Delta n \rangle ^\mathrm{act}_\mathrm{dis} 
= 0.733 \pm 0.013$. 
 
Generating a greater
$\langle \Delta n\rangle_{\rm ind}$
than the passive demon, the active demon generates a greater temperature difference.
Equation~\eqref{eq:somhorsttemperature} defines the temperature $T$ attained by a subset ($A$ or $B$) in one trial. To calculate an average temperature $\langle T \rangle$, we averaged the subset's photon number over trials and unitaries.
We substituted this average photon number and $M=2$ into Eq.~\eqref{eq:somhorsttemperature}.
Figure~\ref{fig:PhNdiff}(c) shows the indistinguishable photons' average temperatures $\langle T \rangle$. Within each pair of bars, the left-hand bar characterizes subset $A$; and the right-hand bar, $B$. The leftmost pair shows the pre-conditional-switching temperatures $T_A^\init$ and $T_B^\init$; and the rightmost pair, the final temperatures $T_A^\final$ and $T_B^\final$. The initial $\langle T \rangle$s lie close together---within one standard deviation of each other. Hence $A$ begins nearly in equilibrium with $B$. $T_A^\final$ lies farther from $T_B^\final$; the conditional switching heightens the average temperatures' separation. Hence the demon drives the photonic system farther from equilibrium.

Particle indistinguishability exacerbates the final state's nonequilibrium nature, we show now.
Figure~\ref{fig:PhNdiff}(a,b) shows the distinguishable-particle and indistinguishable-particle distributions $\mathcal{P}(\Delta n)$. In the passive-demon experiment, the indistinguishable-particle distribution is broad due to bosonic bunching. The distinguishable-particle distribution peaks more sharply. Nevertheless, both distributions are symmetric about 0. Hence 
$\langle \Delta n \rangle _\indist^\mathrm{pas} 
\approx \langle \Delta n \rangle _\dist^\mathrm{pas} \approx 0 \, .$ 

The active demon breaks the symmetry [Fig.~\ref{fig:PhNdiff}(b)]. Bunching leads indistinguishable photons to achieve a greater average photon-number difference than distinguishable photons achieve: 
$\langle \Delta n \rangle _\indist^\mathrm{act} 
> \langle \Delta n \rangle _\dist^\mathrm{act}$. 
This result supports a theoretical prediction in~\cite{Kim_11_Quantum}: Maxwell's demon creates a greater particle-number imbalance from bosons than from distinguishable particles, on average.

We analyze in~\cite{SuppMat} the two primary biases. 
The first arose from our sampling finitely many Haar-random unitaries. The second stemmed from certain detectors' systematically overcounting or undercounting photons. These biases eliminated the Haar-averaged outcome distribution's invariance under mode permutations:
$\langle \Delta n \rangle$ depended on (i) how we partitioned the modes into subsets and (ii) which mode we measured from each subset. These errors impacted $\langle \Delta n \rangle$ negligibly, compared to the active demon's feedback.

\emph{Discussion.---}Using a programmable photonic processor, we have experimentally supported the long-standing prediction that Maxwell's demon benefits from bosonic statistics. We prepared an internally equilibrated system of photons; measured the number of photons in a subset of the modes; and, depending on the outcome, conditionally interchanged the subsets. The process created a difference between two modes' average particle numbers. The average difference reflected how far from equilibrium the whole system evolved. The average difference reached $\langle\Delta n \rangle_\indist = 0.937 \pm 0.027$ if the photons were indistinguishable and $\langle \Delta n \rangle _\dist = 0.733 \pm 0.013$ if they were distinguishable. Hence bosonic statistics boosted the system's evolution away from equilibrium. Modeling the experiment, we extended Ref.~\cite{Kim_11_Quantum}’s theory to arbitrarily many particles and to an experimental preparation procedure. These results experimentally establish a quantum-thermodynamic advantage of bosonic statistics.

This work may find applications in boson-sampling validation, the distinguishing of boson-sampling data from samples produced by efficient classical approximations. According to a no-go theorem, validation is impossible if one can only classically postprocess the samples efficiently~\cite{Hangleiter2019}. 
Only \emph{weak validation} is possible: statistical tests can distinguish samples produced by specific classical simulations. Such tests involve features of the output distribution that are (i) efficiently computable classically and (ii) efficiently estimable from experimental samples. Our experiment provides such a feature: the average photon-number difference, $\langle \Delta n  \rangle$, which photon indistinguishability increases. One can therefore perform a weak-validation test as follows: estimate $\langle \Delta n \rangle$ from the samples. Check whether the value exceeds the distinguishable-photon prediction.

This test benefits from two assets: (a) One can partition the modes into $A$ and $B$ in different ways, obtaining distinct tests. (b) One can effectively partition the modes via postprocessing, after the experiment. These assets benefit adversarial validation, studied recently in the boson-sampling community~\cite{Nikolopoulos2019, Singh2025, Correa2025}. In adversarial validation, the provider supplies samples before the verifier chooses their test. Applying our strategy, the verifier can choose their test---can partition the photons---after receiving the samples. The provider, unable to predict the test, cannot easily engineer the samples to spoof it. Even if the provider guesses the verifier's favorite test, the verifier can run multiple tests.

Beyond boson-sampling validation, this work helps initiate a thermodynamics of linear-quantum-optical systems. They, like traditional thermodynamic systems, are too complex for us to calculate all their properties. We must therefore identify useful efficiently computable properties \cite{Walschaers_2020, VDMeer2021,Phillips2019}. This goal shares its spirit not only with boson-sampling validation, but also with thermodynamics, which highlights heat, work, temperature, and entropy. These macroscopic quantities can illuminate unexplored facets of linear quantum optics, as illustrated by our demon experiment and the thermalization experiment~\cite{somhorst_quantum_2023}. Opportunities for future research include work extraction by a demon. Also, using linear quantum optics, one may observe effects of conserved thermodynamic quantities' failure to commute with each other~\cite{22_Manzano_Non,24_Upadhyaya_Non,23_Majidy_Noncommuting,24_Shahidani_Equilibration}.

\begin{acknowledgements}
N.~Y.~H. thanks Simone Colombo, Edwin Pedrozo-Peñafiel, and Vladan Vuletic for inspirational conversations.
This work received support from the National Science Foundation (QLCI grant OMA-2120757). 
T.~U. acknowledges the support of the Joint Center for Quantum Information and Computer Science through the Lanczos Fellowship, as well as the Natural Sciences and Engineering Research Council of Canada (NSERC), through the Doctoral Postgraduate Scholarship. This publication is part of the Vidi project \textit{At the Quantum Edge}, which is financed by the Dutch Research Council (NWO). This work also received support from HTSM-KIC project \textit{Building Einstein's Dice}. 
\end{acknowledgements}

\onecolumngrid

\bibliography{demonbib}

\begin{thebibliography}{49}
\bibitem [{nis()}]{nist_note}%
  \BibitemOpen
  \href@noop {} {\bibinfo {title} {Certain equipment, instruments, software, or materials are identified in this paper to specify the experimental procedure adequately. such identification is not intended to imply recommendation or endorsement of any product or service by {NIST}; nor is it intended to imply that the materials or equipment identified are necessarily the best available for the purpose.}}\BibitemShut {Stop}%
\end{thebibliography}

\end{document}


\title{Supplementary material: Bosonic statistics enhance Maxwell's demon in photonic experiment} 

\author{Malaquias Correa Anguita}
\affiliation{MESA+ Institute for Nanotechnology, University of Twente, P.O.~Box 217, 7500 AE Enschede, Netherlands}

\author{Sara Marzban}
\affiliation{MESA+ Institute for Nanotechnology, University of Twente, P.O.~Box 217, 7500 AE Enschede, Netherlands}

\author{William F. Braasch Jr.}
\affiliation{Joint Center for Quantum Information and Computer Science, NIST and University of Maryland, College Park, MD 20742, USA}

\author{Twesh Upadhyaya}
\affiliation{Joint Center for Quantum Information and Computer Science, NIST and University of Maryland, College Park, MD 20742, USA}
\affiliation{Department of Physics,
University of Maryland, College Park, MD 20742, USA}

\author{Gabriel Landi}
\affiliation{Department of Physics and Astronomy, University of Rochester, Rochester, New York 14627, USA}

\author{Nicole Yunger Halpern}
\affiliation{Joint Center for Quantum Information and Computer Science, NIST and University of Maryland, College Park, MD 20742, USA}
\affiliation{Institute for Physical Science and Technology, University of Maryland, College Park, Maryland 20742, USA}

\author{Jelmer J. Renema}
\affiliation{MESA+ Institute for Nanotechnology, University of Twente, P.O.~Box 217, 7500 AE Enschede, Netherlands}

\maketitle
\onecolumngrid

These materials are organized as follows. In Supplementary Note~\ref{SM_Analytics}, we calculate the average photon-number difference $\expval{\Delta n}$, assuming that the photons are distinguishable and assuming that they are indistinguishable. We rely on the Haar ensemble's permutation invariance. Supplementary Note~\ref{SM_Exp_setup} describes the experimental setup. In Supplementary Note~\ref{SM_bias_analysis}, we analyze experimental biases that stem from (i) finite sampling of Haar-random unitaries and (ii) systematic detector errors. Supplementary Note~\ref{SM_q_fluxns} describes the role of quantum fluctuations in our demon experiments. Supplementary Note~\ref{SM_construct_Haar} specifies how we constructed Haar-random matrices.

\section{Theoretical calculations of probability distribution over possible photon-number differences}
\label{SM_Analytics}

In this supplementary note, we derive the probabilities plotted in the main text's Fig.~3. The experimental setup generates $N=3$ photons and launches them into an interferometer that has $M=4$ modes. 
Below, we keep $N$ and $M$ arbitrary until Supplementary Note~\ref{sec:photon-diffs}, which we specialize to our setting. 
The $N$ photons enter input modes labeled $i_1, i_2,\ldots,i_N$. Each mode receives, at most, one photon.
Denote by ${\rm U}(M)$ the set of $M\times M$ unitaries.
The interferometer implements a unitary $U\in\mathrm{U}(M)$ on the modes, without disturbing the internal degrees of freedom (e.g., polarizations). 
A detector registers the number $s_j$ of photons in mode $j\in \{1,2,\ldots, M\}$.
An \textit{outcome} is an occupation-number vector $s=(s_1,s_2,\ldots,s_M)$ that satisfies $\sum_{j=1}^M s_j=N$. 
In Supplementary Note~\ref{sec:partial_distinguishability}, we compute the probability of obtaining outcome $s$, given an initial state $|\psi\rangle$.
We compute these probabilities from a fixed, general $U$, assuming the photons are indistinguishable and then assuming the photons are distinguishable.
In Supplementary Note~\ref{sec:averaged-probs}, we average these probabilities over $U$ with respect to the Haar measure on $\mathrm{U}(M)$.
We explain in Supplementary Note~\ref{sec:photon-diffs} how these probabilities lead to the distribution over the difference $\Delta n$ between two output modes' photon numbers. We compute the latter distribution, assuming the demon is passive and assuming it is active.

\subsection{Probability that a fixed interferometer yields a particular outcome}
\label{sec:partial_distinguishability}

In this supplementary note, we first introduce the technical setup. It shows how to model partially distinguishable particles. Next, we review necessary concepts from second quantization, the natural framework for describing multiple quantum particles. We then model the initial multiphoton state, the interferometric transformation, and the measurement. Finally, we compute the outcome probabilities. This material is standard in the theory of multiphoton interference; we follow~\cite{Walschaers_2020}.

\textbf{Technical setup:}
To describe partially distinguishable photons, we employ two Hilbert spaces. 
Denote the spatial-mode space (on which the interferometer enacts a transformation) by $\mathcal{H}_{\rm S} \cong \mathbb{C}^M$; and an orthonormal basis for it, by $\{|e_j\rangle\}_{j=1}^M$.
We need not entirely specify the internal-degree-of-freedom space $\mathcal{H}_{\rm I}$, assumed to be finite-dimensional.
Photons may have position-space wave functions that overlap negligibly; such photons are distinguishable.
Yet if two photons occupy the same spatial mode, they may be indistinguishable, distinguishable, or partially distinguishable.
The overlap between two photons' internal states governs their distinguishability: an overlap of 0 implies distinguishability, whereas an overlap of 1 implies indistinguishability.
We represent pure single-photon states by unit vectors in $\mathcal{H}_{\rm S} \otimes \mathcal{H}_{\rm I}$.
Consider a photon, localized in mode $j$, whose internal degrees of freedom are in the state $|\phi\rangle$. The photon is in the state $|e_j\rangle \otimes |\phi\rangle$.

We now calculate single-photon transition probabilities: the probability that, upon beginning in mode $j$, a photon ends in mode $k$. A unitary $U$ models the interferometer's action on a single-photon spatial-mode state.
Consider injecting into mode $j$ a photon whose internal degree of freedom is in $|\phi\rangle$.
A detector detects the photon in output mode $k$ with a probability  \footnote{The single-photon scenario considered here is simpler than multiphoton scenarios where distinguishability must be considered.} 
\begin{equation}
    p_{j\rightarrow k} =  \big| \left(\langle e_k | \otimes \langle \phi|\right) \left( U\otimes \id\right)\, (|e_j\rangle \otimes |\phi \rangle) \big|^2=|\langle e_k|U|e_j\rangle|^2
    \equiv |U_{kj}|^2 \,.
\end{equation}

\textbf{Tools from second quantization:}
Let $\mathcal{F}(\mathcal{H}_{\rm S}\otimes \mathcal{H}_{\rm I})$ denote the bosonic Fock space over the one-particle Hilbert space $\mathcal{H}_{\rm S}\otimes \mathcal{H}_{\rm I}$.
For every pure state $|\xi\rangle\in \mathcal{H}_{\rm S}\otimes \mathcal{H}_{\rm I}$, we denote the corresponding annihilation operator by $a(\xi)$ and creation operator by $a^\dagger(\xi)$.
They satisfy the canonical commutation relations
\begin{equation}\label{eq:commutation_rel}
   \big[a(e_j\otimes\phi),\,a^\dagger(e_k\otimes\omega)\big]
   =\delta_{j k}\,\langle \phi\mid \omega\rangle\quad \text{and}\quad
   \big[a^\dagger(\bullet),a^\dagger(\bullet)\big]
   =\big[a(\bullet),a(\bullet)\big]=0 \,.
\end{equation}
The vacuum state satisfies $a(\xi)| 0\rangle=0$. We follow the convention according to which $a^\dagger(\bullet)$ is linear (as opposed to antilinear) in its argument: if $c_1, c_2 \in \mathbb{C}$,
\begin{equation}
a^\dagger(c_1\xi+c_2\eta)=c_1\,a^\dagger(\xi)+c_2\,a^\dagger(\eta) \,.
\end{equation}
Let $W$ denote any single-particle unitary defined on $\mathcal H_{\rm S} \otimes\mathcal H_{\rm I}$. 
Second-quantizing it yields a unitary $E(W)$ defined on the Fock space.
This unitary
obeys the covariance property
\begin{equation}\label{eq:E(W)_covariance}
E(W)\,a^\dagger(\xi)=a^\dagger(W\xi) \, E(W) \qquad \,
\end{equation}
and preserves the vacuum:
$E(W) \, | 0\rangle=| 0\rangle$~\cite{Walschaers_2020}.

\textbf{Representation of the interferometer's action:} A $W$ represents the interferometer's action: $W=U\otimes \id$ acts on $\mathcal{H}_{\rm S} \otimes \mathcal{H}_{\rm I}$. 
For the convenience of leveraging the Weingarten calculus below, we express $U$ as a matrix. Furthermore, we express $U$'s action on a creation operator $a^\dagger(\bullet)$ in terms of a matrix representation of $U$.
We derive the expression using the creation operator's linearity, Eq.~\eqref{eq:E(W)_covariance}, and the expansion $Ue_i=\sum_{o=1}^M U_{oi} \,e_o$: 
\begin{equation}
E(U\otimes \id)\,a^\dagger(e_i\otimes\omega)
=a^\dagger\big((U\otimes \id)(e_i\otimes\omega)\big)\, E(U\otimes \id)
=\sum_{o=1}^M U_{oi}\,a^\dagger(e_o\otimes\omega)\, E(U\otimes \id) \,.
\label{eq:E(UxI)}
\end{equation}

\textbf{Representation of the initial state:}
$N$ photons enter distinct input modes labeled $i_1, i_2, \ldots,i_N$. The photons' internal states are $|\psi(1)\rangle,|\psi(2)\rangle,\ldots,|\psi(N)\rangle\ \in \mathcal H_{\rm I}$. The initial state therefore has the form
\begin{equation}\label{eq:initial-state}
|\Psi\rangle=\prod_{k=1}^N a^\dagger\big(e_{i_k}\otimes\psi(k)\big)\,| 0\rangle \,.
\end{equation}

\textbf{Representation of the measurement:}
Detectors resolve each output mode's photon number without measuring the photons' internal states. Consider measuring the set of output modes. We specify the outcome with an occupation-number vector
$s=(s_1, s_2, \ldots,s_M)$ that satisfies $\sum_{j=1}^M s_j=N$. $s_j$ denotes the number of photons detected in mode $j$. We can encode the outcome alternatively in a multiset $O=\{o_1, o_2, \ldots,o_N\}$ in which the $j$-valued element appears $s_j$ times.
For example, $O=\{1,1,4\}$ encodes $s_1=2$, $s_4=1$, and $s_{j \neq 1,2}=0$.\footnote{
Using $O$ does not imply that we can distinguish and label photons.}
Let $d_{\rm I} \coloneqq \dim (\mathcal{H}_{\rm I})$ denote the internal-state space's dimensionality. Consider any orthonormal basis $\{|f_r\rangle\}_{r=1}^{d_{\rm I}}$ of $\mathcal H_{\rm I}$.
In terms of the multiplicities $s_j$, we define the symmetry factor
\begin{equation}
\mu(s)\coloneqq \prod_{j=1}^M s_j! \, .
\end{equation}
We will use this factor to avoid overcounting when $s_j>1$ photons occupy any output mode $j$.

The measurement operator associated with the outcome $O$ (equivalently, with $s$) is
\begin{equation}
P_O=\frac{1}{\mu(s)}\sum_{r_1=1}^{d_{\rm I}}\sum_{r_2=1}^{d_{\rm I}}\ldots\sum_{r_N=1}^{d_{\rm I}}
\left[\prod_{k=1}^N a^\dagger\left(e_{o_k}\otimes f_{r_k}\right)\right]\ketbra{0}{0}
\left[\prod_{\ell=1}^N a\left(e_{o_\ell}\otimes f_{r_\ell}\right)\right] \,.
\label{eq:PO}
\end{equation}
The sums over $r_1, r_2, \ldots,r_N$ reflect the detectors' inability to resolve internal degrees of freedom.
The $1/\mu(s)$ precludes the overcounting associated with permuting identical photons within any mode $j$ that contains $s_j$ photons.

\textbf{Probability calculation:} Consider preparing the photons in a state $|\Psi\rangle$, evolving them under the interferometer, and measuring the number of photons in each output mode. Let us calculate the probability of obtaining outcome $O$,
\begin{equation}
p_{\Psi \rightarrow O}=\langle\Psi|\,E(U^\dagger\otimes I)\,P_O\,E(U\otimes  I)\,|\Psi\rangle \,.
\label{eq:probstart}
\end{equation}
At the end of this subsection, we specialize this probability to distinguishable and to indistinguishable photons. We do so by specifying the overlaps of the internal states in $|\Psi\rangle$ [Eq.~\eqref{eq:initial-state}].

To compute $p_{\Psi \rightarrow O}$, we use the canonical commutation relations~\eqref{eq:commutation_rel} to bring operator products into normal order.
When sandwiched between $\langle 0|$ and $|0\rangle$, all normal-ordered terms vanish; only complex-valued contractions remain.
Concretely, commuting an annihilation operator past a creation operator gives
\begin{equation}
a(e_y\otimes\phi)\,a^\dagger(e_x\otimes\omega)
=a^\dagger(e_x\otimes\omega)\,a(e_y\otimes\phi)+\delta_{y x}\,\langle\phi\mid\omega\rangle \,.
\end{equation}
Repeated application yields the bosonic Wick rule for vacuum expectation values:
\begin{equation}
\langle 0\rvert \left[\prod_{k=1}^n a(e_{x_k}\otimes\phi_k)\right]
\left[\prod_{\ell=1}^n a^\dagger(e_{y_\ell}\otimes\omega_\ell)\right]| 0\rangle
=\sum_{\pi\in S_n}\ \prod_{k=1}^n
\delta_{x_k \, y_{\pi(k)}}\,\langle\phi_k\mid\omega_{\pi(k)}\rangle \,.
\label{eq:wick}
\end{equation}
$S_n$ denotes the symmetric group on $n$ objects (the group of the permutations of $n$ objects). $\pi(k)$ denotes the $k^{\rm th}$ object's image under the permutation $\pi$.

We now compute $p_{\Psi \rightarrow O}$ [Eq.~\eqref{eq:probstart}], in which many instances of the left-hand side of Eq.~\eqref{eq:wick} will appear.
We apply Eq.~\eqref{eq:E(UxI)} and $E(U\otimes \id) |0\rangle = |0\rangle$. The final two factors in Eq.~\eqref{eq:probstart} evaluate to
\begin{align}\label{eq:initial-ket}
   & E(U\otimes\id)\,|\Psi\rangle
=\prod_{k=1}^N\left[\sum_{x=1}^M U_{xi_k}\,a^\dagger\LParen e_x\otimes\psi(k)\RParen\right]| 0\rangle 
=\sum_{x_1=1}^M \sum_{x_2 =1}^M \ldots \sum_{x_N=1}^M\left(\prod_{k=1}^N U_{x_k i_k}\right)\left[\prod_{\ell=1}^N a^\dagger\LParen e_{x_\ell}\otimes\psi(\ell)\RParen\right]| 0\rangle .
\end{align}
The beginning of the $p_{\Psi \rightarrow O}$ expression [Eq.~\eqref{eq:probstart}] evaluates to
\begin{align}
   \label{eq:initial-bra} &
\langle\Psi|\,E(U^\dagger\otimes \id)
=\sum_{x'_1=1}^M \sum_{x'_2 =1}^M \ldots \sum_{x'_N=1}^M\left(\prod_{k=1}^N U^{*}_{x_k' i_k}\right)\langle 0\rvert\left[\prod_{\ell=1}^N a\LParen e_{x_\ell'}\otimes\psi(\ell)\RParen\right]\,.
\end{align}
We insert~\eqref{eq:PO},~\eqref{eq:initial-ket}, and~\eqref{eq:initial-bra} into~\eqref{eq:probstart}.
We then apply the Wick rule~\eqref{eq:wick} to each vacuum correlator, obtaining two sums over permutations in $S_N$.
In Eq.~\eqref{eq:PO}, the sums over the internal basis $\{|f_{r_k}\rangle\}$ yield the completeness relation $\sum_r 
\ketbra{f_r}{f_r}=\id_{\mathcal H_{\rm I}} \, .$ 
Therefore, we can eliminate the internal indices in favor of the overlaps
$\langle \psi(\pi'(k))\mid \psi(\pi(k))\rangle$.
Collecting coefficients yields
\begin{equation}
p_{\Psi\rightarrow O}=\frac{1}{\mu(s)}\sum_{\pi,\pi'\in S_N}
\left(\prod_{k=1}^N U_{o_{\pi(k)} i_k}\,U^{*}_{o_{\pi'(k)} i_k}\right)
\left[\prod_{\ell=1}^N \langle \psi\LParen\pi'(\ell)\RParen\mid \psi\LParen\pi(\ell)\RParen\rangle\right] \,.
\label{eq:general}
\end{equation}
[The last equation slightly generalizes Ref.~\cite{Walschaers_2020}'s Eq.~(142): 
multiple photons can occupy the same output mode in our setup, not in~\cite{Walschaers_2020}.
This generalization leads to the prefactor $1/\mu(s)$.]
If the photons are indistinguishable, the internal-state overlaps equal one.
In this case, the outcome probabilities have the form
\begin{equation}
p^{{\rm (indist)}}_{\Psi\rightarrow O}
= \frac{1}{\mu(s)}
\sum_{\pi, \pi'\in S_N}\prod_{k=1}^N U_{o_{\pi(k)} i_k} {U}_{o_{\pi'(k)} i_k}^* . 
\end{equation}
If the photons are distinguishable, the internal states are orthogonal.
Only $\pi=\pi'$ survives in~\eqref{eq:general}. The probabilities assume the form
\begin{equation}
p^{\mathrm{dist}}_{\Psi\rightarrow O}=\frac{1}{\mu(s)}\sum_{\pi\in S_N}\prod_{k=1}^N \big|U_{o_{\pi(k)} i_k}\big|^2 .
\end{equation}

\subsection{Haar-averaged outcome probabilities}\label{sec:averaged-probs}

We now average $p_{\Psi\rightarrow O}^{\rm indist}$ and $p_{\Psi\rightarrow O}^{\rm dist}$ over the $M\times M$ unitaries $U$ with respect to the Haar measure.
Recall that ${\rm U}(M)$ denotes the group of $M\times M$ unitaries.
Denote the Haar-measure average over ${\rm U}(M)$ by $\mathbb{E}_{{\rm U}(M)}$.
We use identities involving the Weingarten calculus~\cite{Collins_2006, collins_weingarten_2022}, so we first review the concepts required.

\textbf{Unitary Weingarten calculus:}
Let us summarize the unitary Weingarten calculus needed to Haar-average degree-$d$ monomials in unitary-matrix entries.
In our photonics application, we average over ${\rm U}(M)$ to calculate degree-$N$ moments [Haar integrals of products of (i) $N$ matrix entries of $U$ and (ii) $N$ conjugate entries].
Fix a positive integer $d\in\mathbb{Z}_{>0}$; in our application, $d{=}N$ equals the number of photons.
The symmetric group $S_d$ has irreducible representations indexed by partitions $\lambda\vdash d$ of length $\ell(\lambda)$ and character $\chi^\lambda$. 
Let $R\in\mathbb{N}$ denote the dimension of the unitary group ${\rm U}(R)$; in our application, $R{=}M$ equals the number of modes. Define the shifted-content product~\footnote{
This name is justified because $C_\lambda(R)$ is the product, over the boxes $(i,j)$ of $\lambda$'s Young diagram, of a shift of the box \emph{content} $c(i,j)\coloneqq j-i$~\cite[p.~409]{Stanley_Fomin_1999}: $C_\lambda(R)=\prod_{(i,j)\in\lambda}\LParen R+c(i,j)\RParen$. Equivalently, $C_\lambda(R)$ is the numerator in the hook--content formula for the dimension of the $U(R)$ irrep indexed by $\lambda$~\cite[Thm.~7.21.2]{Stanley_Fomin_1999}.} 
(to be used below in a Weingarten-function expansion)
\begin{equation}
    C_\lambda(R)\;\coloneqq\;\prod_{i=1}^{\ell(\lambda)}\ \prod_{j=1}^{\lambda_i} (\,R+j-i\,)\,.
\end{equation}
We can express the \emph{unitary Weingarten function} ${\rm Wg}_{R,d}:S_d\to\mathbb{C}$ in terms of characters:
\begin{equation}
\label{eq:Wg-CMN-2.8}
{\rm Wg}_{R,d}(\sigma)\;=\;\frac{1}{d!}\sum_{\substack{\lambda\vdash d\\ C_\lambda(R)\neq 0}}
\frac{\chi^\lambda(\mathrm{id}_d)\,\chi^\lambda(\sigma)}{C_\lambda(R)}\,.
\end{equation}
This function appears in the identity~\cite{Collins_2006}
\begin{equation}\label{eq:weingarten-ID}
\int_{{\rm U}(R)} 
\left(  U_{i_1 j_1} \ldots U_{i_d j_d}   \right)
\left(  U_{i_1' j_1'}^* \ldots U_{i_d' j_d'}^* \right) dU 
= \sum_{\sigma, \tau \in S_d} 
\left(  \delta_{i_1 \, i_{\sigma(1)}'} \ldots 
\delta_{i_d \, i_{\sigma(d)}'}  \right)
\left(  \delta_{j_1 \, j_{\tau(1)}'} \ldots 
\delta_{j_d \, j_{\tau(d)}'} \right) 
\text{Wg}_{R,d}  \left(  \tau \sigma^{-1}  \right).
\end{equation}
Such integrals appear in the Haar-averaged probabilities
\begin{align}
    & \mathbb{E}_{{\rm U}(M)}\left(p^{{\rm (indist)}}_{\Psi\rightarrow O}\right) 
    = \frac{1}{\mu(s)}
\sum_{\pi, \pi'\in S_N} \int_{{\rm U}(M)} dU \left(\prod_{k=1}^N U_{o_{\pi(k)} i_k} {U}_{o_{\pi'(k)} i_k}^* \right)
    \quad \text{and} \\
    \label{eq:average-distinguishable}
    & \mathbb{E}_{{\rm U}(M)}\left(p^{{\rm (dist)}}_{\Psi\rightarrow O}\right) 
    =
    \frac{1}{\mu(s)}\sum_{\pi\in S_N}
    \int_{{\rm U}(M)} dU\left(
    \prod_{k=1}^N \big|U_{o_{\pi(k)} i_k}\big|^2
    \right) .
\end{align}
We also use the following summed Weingarten identity,
\begin{equation}\label{eq:wg-plain-sum}
    \sum_{\sigma \in S_d} {\rm Wg}_{R,d}(\sigma) = \frac{1}{R (R+1) (R+2) \ldots (R+d - 1)}
\end{equation}
(e.g., Example 3.1 of~\cite{collins_integration_2014} and Prop.~45 of~\cite{kostenberger_weingarten_2021}).

\textbf{Indistinguishable photons:}
If the photons are indistinguishable, the Haar-averaged outcome probability has the form
\begin{align}
    \mathbb{E}_{{\rm U}(M)}\left(p^{{\rm (indist)}}_{\Psi\rightarrow O}\right) 
    &=
    \frac{1}{\mu(s)}
    \sum_{\pi, \pi'\in S_N} \int_{{\rm U}(M)} dU \left(\prod_{k=1}^N U_{o_{\pi(k)} i_k} {U}_{o_{\pi'(k)} i_k}^* \right)\\
    &= \frac{1}{\mu(s)} \sum_{\pi, \pi'\in S_N}
    \sum_{\sigma,\tau\in S_N}
    \left(\prod_{k=1}^N \delta_{o_{\pi(k)}  o_{\pi'\LParen\sigma(k)\RParen}} \delta_{i_k i_{\tau(k)}}\right) 
    {\rm Wg}_{M,N} \left( \tau \sigma^{-1} \right)
    \\
    &= \frac{1}{\mu(s)} \sum_{\pi, \pi'\in S_N}
    \sum_{\sigma\in S_N}
    \left(\prod_{k=1}^N 
    \delta_{o_{\pi(k)} o_{\pi'\LParen\sigma(k)\RParen}} \right) 
    {\rm Wg}_{M,N} \left( \sigma^{-1} \right) .
\end{align}
The second equality required Eq.~\eqref{eq:weingarten-ID}.
The third equality follows from the input photons' occupying distinct modes, such that the $\delta_{i_k i_{\tau(k)}}$ implies $\tau = \id$.
To continue, we consider fixing the permutations $\sigma$ and $\pi'$ in the last equation.
Applying $\delta_{o_{\pi(k)} o_{\pi'\LParen\sigma(k)\RParen}}$, for all $k$, is equivalent to saying that the list $(o_{\pi(1)}, o_{\pi(2)}, \ldots, o_{\pi(N)})$ equals $(o_{\pi'\LParen\sigma(1)\RParen}, o_{\pi'\LParen\sigma(2)\RParen}, \ldots, o_{\pi'\LParen\sigma(N)\RParen})$.
Of the $N!$ permutations $\pi$, exactly $\mu(s)$ of them satisfy this constraint.
Therefore, the sum over $\pi$ and $\pi'$ evaluates to $N!\, \mu(s)$. Thus,
\begin{equation}
    \mathbb{E}_{{\rm U}(M)}\left(p^{{\rm (indist)}}_{\Psi\rightarrow O}\right) = N! \sum_{\sigma \in S_N} {\rm Wg}_{M,N}\left(\sigma^{-1}\right) .
\end{equation}
The identity~\eqref{eq:wg-plain-sum} implies
\begin{align}
    \mathbb{E}_{{\rm U}(M)}\left(p^{{\rm (indist)}}_{\Psi\rightarrow O}\right) 
    &=\frac{N!}{M (M+1) (M+2) \ldots (M+N - 1)}
    = {1}\bigg/{\binom{M+N-1}{N}} .
\end{align}
If the photons are indistinguishable, the Haar-averaged probability is uniform over the final occupancy patterns $O$ that feature $N$ photons distributed across $M$ modes. The patterns number $\binom{M+N-1}{N}$.

\textbf{Distinguishable photons:}
The Haar-averaged outcome probabilities do not simplify as neatly if the photons are distinguishable as if they are indistinguishable.
Applying Eq.~\eqref{eq:weingarten-ID} to Eq.~\eqref{eq:average-distinguishable} yields
\begin{equation}\label{eq:average-distinguishable_2}
    \mathbb{E}_{{\rm U}(M)}\left(p^{{\rm (dist)}}_{\Psi\rightarrow O}\right) 
    =
    \frac{1}{\mu(s)}\sum_{\pi \in S_N} \sum_{\sigma\in S_N} \left(
    \prod_{k=1}^N \delta_{o_{\pi(k)} o_{\pi(\sigma(k))}}
    \right) {\rm Wg}_{M,N}\left(\sigma^{-1}\right) .
\end{equation}
Reference~\cite{photonic_maxwells_demon_analytics_v0_1_0} shows our evaluation of this expression at $(M,N) = (4,3)$.

\subsection{Computation of the probability distribution over the possible photon-number differences}
\label{sec:photon-diffs}

We now apply the multiphoton-interference probabilities to our demon experiment: we compute the distribution over $\Delta n \coloneqq n_A - n_B$ over the imbalance between one $A$ mode's photon number, $n_A$, and one $B$ mode's, $n_B$. We assume that the demon is active, then assume that he is passive. The results appear in the main text's Fig.~3 as the bars, labeled ``analytical.'' 

Let us apply the general results above to $M=4$ and $N=3$.
By the modes' permutation symmetry, we may label any two modes as subset $A$, labeling the complementary modes as $B$.
We choose $A = \{1, 2\}$ and $B=\{3, 4\}$, as in the main text.
Similarly, we may compute the single-mode photon 
number $n_A$ from any $A$ mode and compute the analogous $n_B$ from any $B$ mode. We choose $n_A = s_1$ and $n_B = s_3$ (recall that $s_j$ denotes the number of output photons in mode $j$).

Recall that $\Delta n \coloneqq n_A - n_B$ equals the number of output photons in mode 1, minus the number in mode 3. We compute the probability distribution over the $\Delta n$ values.
First, we compute $\Delta n$ for each detection outcome $O$. 
We then sum the probabilities $p_{\Psi\rightarrow O}^{\rm(dist/indist)}$ associated with the same $\Delta n$. These steps yield the probability distribution if the demon is passive.
Suppose that the demon is active. For each detection outcome $O$, we interchange modes 1 and 3 if and only if $s_1 + s_2 < s_ 3 + s_4$.
Then, we compute $\Delta n$ (using the modes' new labelings, if we interchanged modes). Finally, we sum $p_{\Psi\rightarrow O}^{\rm(dist/indist)}$ over the outcomes that yield the same $\Delta n$. 
We implement this procedure in the code in~\cite{photonic_maxwells_demon_analytics_v0_1_0}.

\section{Experimental setup}
\label{SM_Exp_setup}

\begin{figure*}[h!]
    \includegraphics[width = 0.95\textwidth]{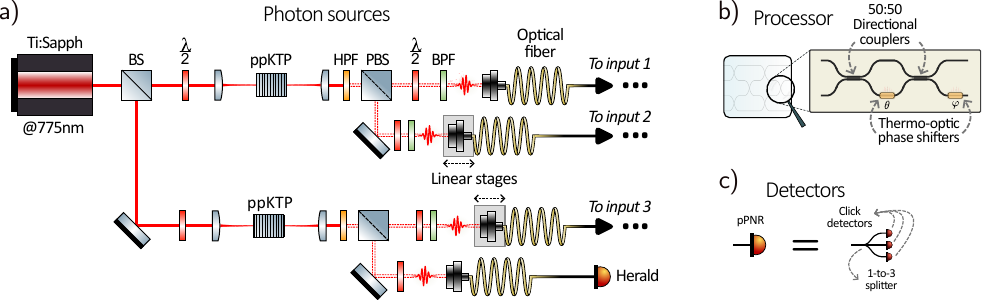}
    \caption{
    {(a) Photon sources:} A 775 nm, Ti:Sapph laser pumps two Type II ppKTP crystals. A high-pass filter (HPF) removes residual pump light before a polarizing beam splitter (PBS) splits photon pairs. Then, the photons are coupled into optical fibers. Three photons enter the processor, while a fourth serves as a herald.
    {(b) Photonic processor:} Detailed schematic of the chip’s actuation mechanism. Adjacent waveguides approach each other and separate, coupling together via their evanescent fields. Thermo-optical phase shifters, situated in one waveguide, enable tunable gates. Each gate acts on two optical modes as a $2 \times 2$ unitary matrix. These gates form a checkerboard pattern, enabling fully reconfigurable $M \times M$ unitary matrices.
    {(c) Detectors:} Three superconducting-nanowire single-photon detectors, together with a one-to-three splitter, implement a pseudo-photon-number-resolving (pPNR) measurement. 
    }
    \label{fig:app-setup}
\end{figure*}

Figure~\ref{fig:app-setup} shows a schematic of the experiment. The photon source consists of two periodically poled potassium titanyl phosphate (ppKTP) crystals in a Type II degenerate configuration. The crystals spontaneously convert a pump photon at 775\,nm into a pair of photons at 1550\,nm. We filtered the photons to a bandwidth of approximately 12\,nm to enhance the two-photon state’s purity.

We can continuously control the photons' degree of distinguishability by adjusting the times at which the photons arrive at the chip. To quantify this distinguishability, we performed on-chip Hong–Ou–Mandel (HOM) dip experiments. They imply a lower bound on the overlap between the input photons' states. To define the overlap, we denote photon $j$'s state by $\ket{\psi_j}$.
Photons $j$ and $k$ have a state overlap 
$x_{j,k} \coloneqq \left| \langle \psi_j | \psi_k \rangle \right|$. The HOM-dip visibility $V$ bounds the overlap as $V \leq (x_{j,k})^{2}$. We measured visibilities of 90\,\% and 95\,\%, using photons from different sources, and 98\,\%, using photons from the same source. 

The photonic processor is a 12-mode ($M {=} 12$) universal interferometer constructed from silicon-nitride waveguides. The optical transmission ranges from 2.2\,dB to 2.7\,dB. (The value depends on the optical path \cite{Triplex, QuiX2021}.) 
The processor implements arbitrary optical transformations from $M$ input to $M$ output modes, preserving the photons' second-order coherence \cite{QuiX2021}. Let $U_{\rm tar}$ denote a target transformation; and $U_{\rm exp}$, the transformation implemented experimentally.
We quantify the similarity between $U_{\rm tar}$ and $U_{\rm exp}$ with the amplitude fidelity 
$F \coloneqq \frac{1}{M}\mathrm{Tr}(|U^{\dagger}_{\rm tar}| \cdot | U_{\rm exp}|)$. We measured $F = 0.98$, on average over experiments.

We conducted thermodynamic simulations using various chip configurations. An interferometer transformed a subset of the 12 modes, or 2 interferometers transformed subsets sequentially [Figs.~1(c,d) of the main text]. We realized these configurations by selectively activating regions of the photonic processor. For instance, a 3-mode transformation involved the top left-most $3\times 3$ unit cells. A 4-mode transformation involved a $4\times 4$ region. When concatenating transformations, we implemented one on the chip's top left-hand corner and the other transformation on the top right-hand corner. This spatial separation reduced crosstalk to a negligible level.

We measured the photonic system's output using 13 superconducting-nanowire single-photon detectors \cite{ReviewSNSPD, Marsili-SNSPD}. Twelve detectors monitored the the chip's output. The thirteenth detector served as a herald. We combined the herald with postselection on threefold coincidences at the output. Via this strategy, we effectively prepared an $N{=}3$ boson-sampling input state: one photon in each of the first three modes~\cite{tillmann_2013}. Since we used only the processor’s top four modes ($M = 4$), we multiplexed each output mode to three detectors. We connected each output mode to a one-to-three splitter. Each of the three fibers emerging from the splitter was routed to a separate detector. This multiplexing enabled pseudo-photon-number resolution (pPNR)~\cite{feito_measuring_2009}. 
Across trials, pPNR reconstructs the average-over-trials number of photons per mode. 

We collected data about four-photon events (featuring three photons in the processor, plus the herald photon) at a rate of 1.5\,Hz. These events served as samples of the experiment’s possible outcomes. From the events' frequencies, we estimated the probability distribution attributable to the corresponding experimental configuration. We divided the experiments into two categories, characterized  by the interferometer size: (i) experiments involving only a 3-mode interferometer [Fig.~1(b) of the main text] and (ii) experiments involving at least one 4-mode interferometer [Figs.~1(b,c) and~3 of the main text]. Larger interferometers entail larger output spaces. Therefore, we tailored our data-acquisition times to achieve comparable uncertainties in all the estimated probability distributions. We acquired data for approximately 7 minutes per Haar-random unitary in the type-(i) experiment; and for approximately 15 minutes per unitary, in each type-(ii) experiment.

\section{Analysis and reduction of mode-dependent bias}
\label{SM_bias_analysis} 

\definecolor{dist_blue}{RGB}{0, 81, 199}
\definecolor{indist_red}{RGB}{ 255, 52, 52}

In the demon experiments, we averaged $\Delta n \coloneqq n_A - n_B$ over the 100 measured Haar-random matrices. The average photon-number difference $\langle \Delta n \rangle$ deviated slightly from zero. This deviation was unexpected: the Haar ensemble is symmetric with respect to all permutations of the modes. The ideal distribution $\mathcal{P}(\Delta n)$ shares this symmetry, so $\langle \Delta n \rangle = 0$, ideally.
(See the main-text section ``Photonic Maxwell demon enhanced by particle-number statistics.'') The bias is particularly noticeable in the passive-demon experiment's results [Fig.~3(a) of the main text]. These results followed from partitioning the output modes as $A = \{1,2\}$ and $B = \{3,4\}$. 
The bias stems partially from finite-sampling effects: we averaged over finitely many (100) Haar-random matrices. However, additional experimental imperfections likely contribute. We identify other sources of the bias in Supplementary Note~\ref{SM_bias_partition}. In Supplementary Note~\ref{SM_avg_bias}, we introduce a mitigation strategy based on mode randomization. The randomization confirms our results rigorously but is unnecessary for observing the Haar-averaged particle-number imbalance engineered by the active demon, we confirm in Supplementary Note~\ref{SM_bias_fix_part}. Due to this confirmation, the main-text results followed from a fixed partition, rather than from the randomization scheme.

\subsection{Bias that stems from choice of partition}
\label{SM_bias_partition}

Tables~\ref{tab:app-PhNdiff_originalPartition} present the photon-number differences $\langle \Delta n \rangle$ averaged over trials and unitaries. The top table reports on the passive-demon experiment; and the bottom table, on the active-demon experiment. Within each table, the top line reports on experimental data; and the bottom line, on numerical simulations of the ideal experiment (which excludes experimental noise). The blue, lefthand data describe distinguishable particles; and the red, right-hand data, indistinguishable particles. Each uncertainty represents the standard error of the mean across the 100 Haar-random unitaries.

\begin{table}[h!]
    \centering
    \begin{tabular}{c||c|c}
    \multicolumn{3}{c}{\large Passive demon}\vspace{0.15cm}\\
         & Distinguishable & Indistinguishable \\
         \hline
         Experiment & \textcolor{dist_blue}{$-0.070\pm0.018$} & \textcolor{indist_red}{$-0.082\pm0.029$} \\
         Simulation & \textcolor{dist_blue}{$-0.055\pm0.018$} & \textcolor{indist_red}{$-0.055\pm0.030$}\vspace{0.5cm} \\
    \multicolumn{3}{c}{\large Active demon}\vspace{0.15cm}\\
         & Distinguishable & Indistinguishable \\
         \hline
         Experiment & \textcolor{dist_blue}{$0.733\pm0.013$} & \textcolor{indist_red}{$0.937\pm0.027$} \\
         Simulation & \textcolor{dist_blue}{$0.701\pm0.012$} & \textcolor{indist_red}{$0.877\pm0.028$} \\
    \end{tabular}
    \caption{Average photon-number differences $\langle \Delta n \rangle$ obtained from 100 Haar-random unitaries. The output modes were partitioned into $A=\{1,2\}$ and $B=\{3,4\}$. See the text for a description of the tables.
     }
    \label{tab:app-PhNdiff_originalPartition}
\end{table}

The passive-demon table merits contrasting with the ideal passive-demon result, which we describe now. Imagine averaging $\Delta n$ over infinitely many Haar-random unitaries. The Haar ensemble exhibits permutation invariance: symmetry with respect to interchangings of the output modes. This permutation invariance would imply that the passive demon's $\langle \Delta n \rangle = 0$.

In contrast, the passive-demon table exhibits the following patterns. The average photon-number difference $\langle \Delta n \rangle$ always assumes a negative, small-magnitude value. The experimental and numerical averages deviate slightly from zero. Therefore, experimental imperfections cannot fully explain the experimental $\langle \Delta n \rangle$'s deviation from the ideal value of 0. The deviation must arise partially from a nonideality common to the experiment and numerics. Finite sampling is the obvious such nonideality.

These patterns contrast with those exhibited by the active-demon table. The $|\langle \Delta n \rangle|$ values there are substantially larger than the passive-demon ones. Regardless of the particles' distinguishability, the experimental and numerical values do not agree to within their uncertainties. This disagreement suggests that finite-sampling effects (common to the experiment and numerics) do not fully explain the active demon's (experimental and numerical) offsets $\langle \Delta n \rangle \neq 0$. Rather, the data suggest a systematic bias in the photodetection.

\begin{figure}[t]
    \centering
    \includegraphics[width=0.9\linewidth]{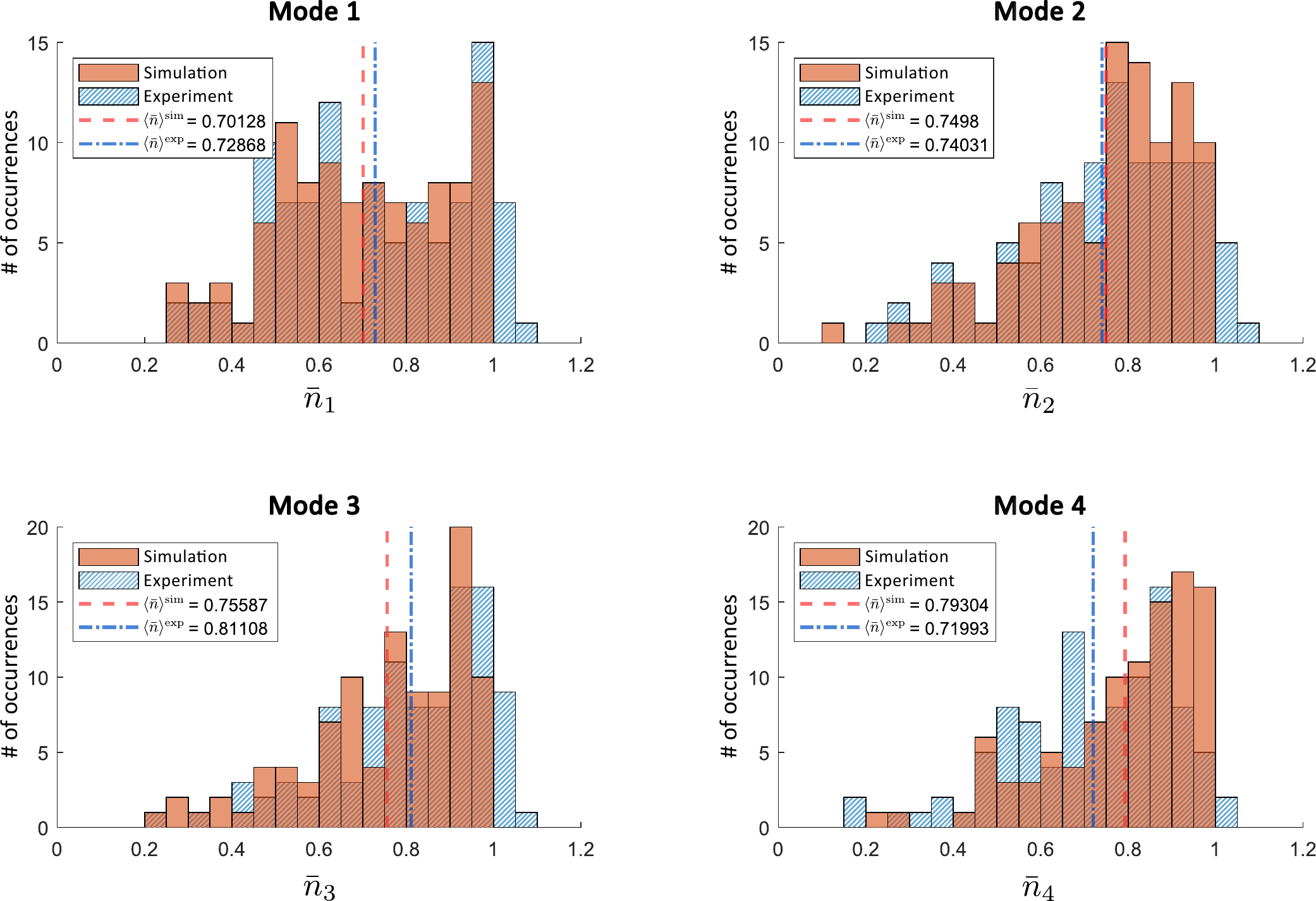}
    \caption{Histogram of the photon number $\bar{n}_j$ per mode, averaged over same-unitary trials. The mode ($j = 1,2,3,4$) varies from plot to plot. Each histogram resulted from computing one $\bar{n}_j$ value per Haar-random unitary, then binning the resulting values. All data describe indistinguishable particles.
    Vertical lines show the values $\langle \bar{n}_j \rangle$ averaged over trials and over the 100 unitaries.
    Orange elements (solid bars and dashed lines) depict numerical simulation data. Blue elements (striped bars and dot–dash lines) depict experimental data.
    }
    \label{fig:app-PhFluxes}
\end{figure}

Figure~\ref{fig:app-PhFluxes} contains the key to understanding this bias. Recall that different detectors measured different output modes' photon numbers. The figure shows each output mode's photon-counting statistics. For every mode ($j = 1,2,3,4$), we computed the average-over-trials photon number $\bar{n}_j$ separately for each Haar-random unitary. We histogrammed the resulting 100 $\bar{n}_j$ values. The solid orange bars follow from numerical simulations; and the striped blue bars, from experiments. The vertical red dashed lines result from averaging the numerical $\bar{n}_j$ values over the 100 Haar-random unitaries. The blue dot-dashed vertical lines show the experimental $\langle \bar{n}_j \rangle$ values. All data represent indistinguishable photons. 

Three observations stand out:
\begin{enumerate}
    \item We preface the first observation with a idealized prediction. Suppose that (i) the modes were ideal and (ii) we averaged over infinitely many Haar-random unitaries. Due to the Haar ensemble's permutation invariance, the modes would exhibit identical statistics.
    In an ($N{=}3$, $M{=}4$) experiment, each mode would contain $N/M = 3/4$ photons, on average. This prediction does not quite match the observed ensemble-averaged photon numbers $\langle \bar{n}_j \rangle$. (Dashed vertical lines represent numerical ensemble averages; and dot-dashed lines, experimental ones.) The simulations are of noiseless experiments. Therefore, experimental noise alone cannot underlie the experimental $\langle \bar{n}_j \rangle \neq 3/4$ values, which the numerics approximately reproduce. A nonideality common to the numerics and experiments must contribute. The finite sampling size (the finite number of Haar-random matrices used) therefore appears to partially explain the deviations from $\langle \bar{n}_j \rangle = 3/4$.
    
    \item The experimental values deviate from the numerical simulation values. This deviation is consistent with experimental imperfections, such as (i) partial photon distinguishability and (ii) the inaccuracy in the chip's implementation of the target unitary.
    
    \item Ideally, each output mode's average photon number is upper-bounded by 1; violations of this bound signal experimental imperfections. Let us derive the bound. A matrix represents the interferometer unitary $U$ relative to the Fock basis; let $U_{ij}$ denote the $(i, j)$ element. Let $n_i^{\mathrm{in}}$ denote the number of photons in input mode $i$. The average photon number in output mode $j$ is 
    $\bar{n}_j 
     = \sum_i |U_{ij}|^2 \, n_i^{\mathrm{in}}$~\footnote{
     Let $\ket{\Psi_{\rm in}}$ denote the input state, $a_j$ denote the creation operator associated with input mode $j$, and $b_k$ denote the creation operator associated with output mode $k$. In the Heisenberg picture, output mode $j$ has an average-over-trials photon number
     $\bar{n}_j = \bra{\Psi_{\rm in}} b_j^\dag b_j
     \ket{ \Psi_{\rm in} }$. 
     The unitary transforms input to output modes as
     $b_j = \sum_k U_{jk} a_k$. Hence
     $\bar{n}_j = \bra{ \Psi_{\rm in} }
     \left( \sum_k U_{jk} a_k \right)^\dag
     \left( \sum_\ell U_{j \ell} a_\ell \right)
     \ket{ \Psi_{\rm in} }
     = \sum_{k, \ell} U_{kj}^*  U_{j \ell}
     \bra{ \Psi_{\rm in} }  a_k^\dag a_\ell
     \ket{ \Psi_{\rm in} }$.
     The inner product is nonzero if and only if $\ell = k$, because annihilation operators annihilate the vacuum state. The inner product becomes $\delta_{k \ell} \, ,$ and
     $\bar{n}_j = \sum_k | U_{kj} |^2$.
     }.
    Each input mode $i$ contains $n_i^{\mathrm{in}} \leq 1$ photon. Therefore, 
    $\bar{n}_j = \sum_i |U_{ij}|^2$.
    Unitarity implies that
    $\sum_i |U_{ij}|^2 = 1$.
    Hence the average output-mode-$j$ photon number ideally satisfies 
    \begin{align}
       \label{eq_n_j_bd}
       \bar{n}_j \le 1 .
    \end{align}
    In the passive-demon experiment, a small number of experimental estimates of the average output-mode photon number $\bar n_j$ exceed the bound in Eq.~\eqref{eq_n_j_bd}, as evidenced by the populated $\bar n_j>1$ bins in the experimental (blue, striped) histograms of Fig.~\ref{fig:app-PhFluxes}. The violations indicate that the experiment breaks at least one of the two assumptions behind Eq.~\eqref{eq_n_j_bd}: (i) each input mode contains $\leq 1$ photon, and (ii) the evolution is unitary. The two experimental imperfections are the most likely culprits:
    \begin{enumerate}[label=(\roman*)]
        \item The source has a nonzero (small) probability of emitting more than one photon pair within an experimental time window. Such multipair emissions ordinarily fail the four-photon postselection. They can contribute to the postselected data, however, if combined with photon loss. Such contributions would violate assumption (i) above.
        
        \item Small miscalibrations of relative detector efficiencies, propagated through the pPNR reconstruction, would introduce nonunitary noise.
    \end{enumerate}
 
\end{enumerate}

We now summarize the mode-dependent bias's consequences for the demon analysis. The fixed-unitary, average-over-trials photon numbers $\bar{n}_j$ deviate from the Haar-ensemble prediction $N/M = 3/4$. Such deviations cannot be uniform across modes, because the interferometer conserves the total photon number. Hence the deviations signal a departure from the permutation invariance characteristic of the ideal Haar average. The departure from permutation invariance introduces systematic biases into analyses based on the modes' statistical equivalence, such as the experimental inference of $\langle \Delta n \rangle$. Modes 3 and 4 exhibit (i) the greatest discrepancies between their experimental and numerical ensemble-averaged averages $\langle \bar{n}_j \rangle$ and (ii) the greatest departures from $\langle \bar{n}_j \rangle=3/4$. Hence the partition $(A, B) = (\{1,2\}, \{3,4\})$ accentuates the asymmetry in the active demon's experimental $\langle \Delta n \rangle$ values.

\subsection{Averaging out mode-dependent bias}
\label{SM_avg_bias}

The setup's noise sources are layered an interdependent; hence correcting every imperfection in the experimental apparatus would be impractical. Instead, we adopt a pragmatic approach to estimating the ideally unbiased $\langle \Delta n \rangle$ (average-over-unitaries, average-over-trials photon-number difference) induced by the demon. We introduce a randomization protocol that averages out the mode-dependent bias. The randomization reduces by over an order of magnitude the passive demon's $\langle \Delta n\rangle$, which ideally vanishes. If the demon is active, the randomization preserves $\langle \Delta n \rangle$'s positive nature and large magnitude. This randomized $\langle \Delta n \rangle$ serves as an estimate of the active demon's ideally unbiased average particle-number imbalance. 

The randomization protocol works as follows. For each Haar-random unitary, we assign the four output modes to the subsets ($A$ and $B$) according to a random partition. Also, we randomly select one mode from each subset; we measure the number of photons in this mode. Using classical postprocessing, we reprocess the recorded data according to the partition, effecting the demon's switching. We compute the $\Delta n$ associated with the unitary. Performing this procedure with each of the 100 Haar-random unitaries, we complete one round of the randomization protocol. We average $\Delta n$ over the Haar-random unitaries. This average provides an estimate of the ideally unbiased demon-induced average photon-number imbalance, regardless of whether the demon is passive or active.

Table~\ref{tab:random} shows the ensemble-averaged photon-number differences $\langle \Delta n \rangle$ obtained from the original mode partition and from the random partition. We have averaged the random-partition results over 1\,000 rounds of the randomization protocol. The uncertainty equals one standard deviation of the mean over the ensemble average.


\begin{table}[h!]
    \centering
    \begin{tabular}{c||c|c}
    \multicolumn{3}{c}{\large Passive demon}\vspace{0.15cm}\\
         Partition type & Distinguishable & Indistinguishable \\
         \hline
         Original   & \textcolor{dist_blue}{$-0.070\pm0.018$} & \textcolor{indist_red}{$-0.082\pm0.029$} \\
         Randomized & \textcolor{dist_blue}{$-0.0007\pm0.0314$} & \textcolor{indist_red}{$-0.0012\pm0.0307$}\vspace{0.5cm} \\
    \multicolumn{3}{c}{\large Active demon}\vspace{0.15cm}\\
         Partition type & Distinguishable & Indistinguishable \\
         \hline
         Original   & \textcolor{dist_blue}{$0.733\pm0.013$} & \textcolor{indist_red}{$0.937\pm0.027$} \\
         Randomized & \textcolor{dist_blue}{$0.7095\pm0.0112$} & \textcolor{indist_red}{$0.8906\pm0.0225$} \\
    \end{tabular}
    \caption{
    Ensemble-averaged photon-number differences $\langle \Delta n \rangle$, obtained using the original mode partition and the randomization protocol. See the text for details.
    }
    \label{tab:random}
\end{table}

The table exhibits the following behaviors. First, consider the passive demon. The randomization substantially decreases the ensemble-averaged photon-number differences's magnitudes, $| \langle \Delta n \rangle |$. If the photons are distinguishable, the randomization drops $| \langle \Delta n \rangle |$ from $0.07$ to $0.0007$. If the photons are indistinguishable, $| \langle \Delta n \rangle |$ drops from $0.082$ to $0.0012$. In both cases, $| \langle \Delta n \rangle |$ decreases by approximately two orders of magnitude. These results indicate the randomization's effectiveness. The randomization moves $\langle \Delta n \rangle$ closer to the 0 expected from the Haar ensemble's permutation invariance. Hence the randomization reduces the mode-dependent bias.

In contrast, the randomization scarcely affects the active-demon results. If the photons are distinguishable, the randomization slightly decreases $\langle \Delta n \rangle$ from 0.733 to 0.710. If the photons are indistinguishable, $\langle \Delta n \rangle$ decreases from 0.937 to 0.891. Hence indistinguishable photons produce a greater $\langle \Delta n \rangle$ than do distinguishable photons. The enhancement is 28 \% if we use the original partition and is 25 \% after randomization. We interpret the 25 \% enhancement as the ideally unbiased $\langle \Delta n \rangle$ induced by the photons' indistinguishability in the active-demon experiment.

\subsection{The active demon induces a significant Haar-averaged particle-number imbalance, even if the mode configuration is fixed.}
\label{SM_bias_fix_part}

The randomization protocol introduced variability into the mode assignments, to average out biases. This subsection features the opposite approach: we analyze the data using fixed mode configurations. Below, we describe our methodology and then analyze its implications. According to this exhaustive approach, indistinguishable photons produce a greater $\langle \Delta n \rangle$ than distinguishable photons do, even in the most biased configurations.

We analyze the data as follows. First, we define a configuration as (i) a partition of the four output modes into subsets $A$ and $B$ and (ii) a choice of one to-be-analyzed mode per subset. Twenty-four configurations exist: we can partition the four modes in ${4 \choose 2} = 6$ distinct ways; and, given a partition, we can choose the analyzed modes in $2 \times 2 = 4$ ways. Using each possible configuration, we process the data from the passive- and active-demon experiments, analyzing the distinguishable and indistinguishable photons.

\begin{figure}[H]
    \centering
    \includegraphics[width=0.75\linewidth]{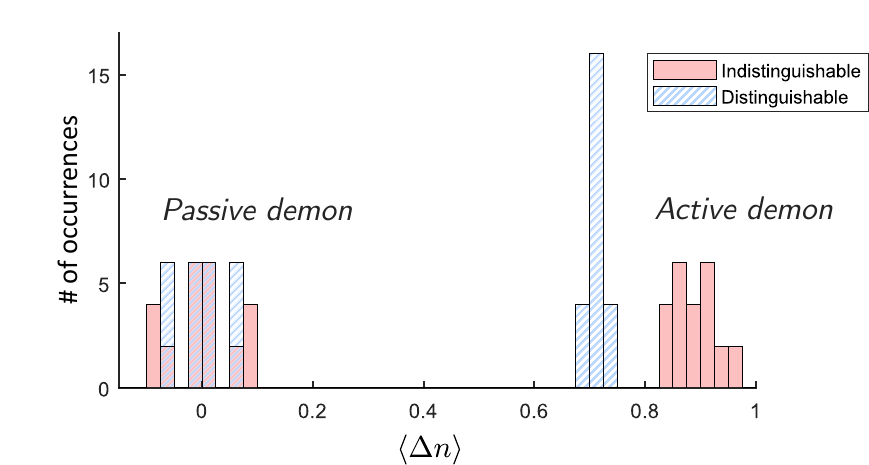}
    \caption{
    Histograms of average-over-trial, average-over-unitary photon-number differences $\langle \Delta n \rangle$ generated from all 24 mode configurations. See the text for details.
    }
    \label{fig:app-everyModeChoice}
\end{figure}

Figure~\ref{fig:app-everyModeChoice} shows histograms of the Haar-averaged photon-number differences $\langle \Delta n \rangle$ computed from experimental data and all 24 configurations. 
Solid red bars represent indistinguishable photons; and striped blue bars, distinguishable photons. The left-hand cluster of bars represents the passive demon; and the right-hand cluster, the active demon. 

The histograms exhibit the following behaviors. First, consider the passive-demon results (the left-hand cluster). The $\langle \Delta n \rangle$ values are centered near zero. However, they vary across configurations. This spread reflects $\langle \Delta n\rangle$'s mode dependence, revealed when we fix the partition, rather than averaging over the possible partitions. 

Now, consider the active-demon results (the right-hand cluster of bars). All the $\langle \Delta n \rangle$ values are positive. The distinguishable-photon data (blue stripes) are clearly separated from the indistinguishable-photon data (solid red bars). Furthermore, all the indistinguishable-photon $\langle \Delta n \rangle$ values exceed all the distinguishable-photon ones. This ordering's consistency demonstrates that the active demon persistently prepares indistinguishable photons far from equilibrium.

\section{Role of quantum fluctuations in our demon experiments}
\label{SM_q_fluxns}

In the main text, we claim that quantum fluctuations dominate over classical fluctuations in our experiment. We support the claim here. By \emph{fluctuations}, we mean run-to-run variations in measurement outcomes. Fluctuations can arise from randomness in state preparations, evolutions, or measurements. Only measurements are random if the system is quantum, closed, prepared in a pure state, and evolved unitarily. Otherwise, fluctuations can arise from classical sources: couplings to environments, noise in control parameters, mixed-state preparation, etc. In our experiment, control-parameter noise and mixed-state preparations lead primarily to static imperfections, rather than to run-to-run fluctuations. In contrast, environmental couplings introduce stochastic photon loss into unobserved modes. Photon loss therefore poses the greatest risk of classical fluctuations. 

We suppressed photon loss by postselecting on trials in which photodetectors registered exactly three output photons. Such postselection could effectively render the dynamics unitary, under two conditions~\cite{Carolan2015}: (i) The single-photon transmission probabilities were approximately uniform across the input–output mode pairs~\cite{clements_optimal_2016}. Our experiment approximately satisfied this condition~\cite{QuiX2021}. (ii) Multipair emissions were sufficiently rare: even when followed by loss, they contributed to the postselected data negligibly. Indeed, the multipair-emission probability was less than one percent of the single-pair-emission probability. Furthermore, loss suppressed multipair emissions' contributions to the postselected data. Hence the photonic system remained approximately closed. Quantum measurements dominated the run-to-run fluctuations in measurement outcomes.

\section{Construction of Haar-random unitaries}
\label{SM_construct_Haar}

We generated Haar-random unitaries by following a standard numerical procedure based on QR decomposition. We first constructed an $M \times M$ complex matrix $A$ whose entries were independent, identically distributed complex Gaussian random variables. We drew the real and imaginary parts from independent normal distributions. Then, we performed a QR decomposition of the matrix: $A = QR$. $Q$ is unitary, and $R$ is upper-triangular. $Q$ serves as a Haar-random unitary used in our experiments and simulations. This procedure samples unitary matrices according to the Haar measure on ${\rm U}(M)$; it is used widely in numerical studies of random unitaries~\cite{mezzadri_how_2007}.

\nocite{Stanley_Fomin_1999}

\bibliography{demonbib}